%% file: ASIA-CCS-sigconf.tex
\documentclass[sigconf]{acmart}
\usepackage{balance}
\def\BibTeX{{\rm B\kern-.05em{\sc i\kern-.025em b}\kern-.08emT\kern-.1667em\lower.7ex\hbox{E}\kern-.125emX}}
    
\copyrightyear{2020} 
\acmYear{2020} 
\setcopyright{acmcopyright}
\acmConference[UMAP '20]{Proceedings of the 28th ACM Conference on User Modeling, Adaptation and Personalization}{July 14--17, 2020}{Genoa, Italy}
\acmBooktitle{Proceedings of the 28th ACM Conference on User Modeling, Adaptation and Personalization (UMAP '20), July 14--17, 2020, Genoa, Italy}
\acmPrice{15.00}
\acmDOI{10.1145/3340631.3394860}
\acmISBN{978-1-4503-6861-2/20/07}

\usepackage{graphicx}  %Required
\usepackage{booktabs} % For formal tables
\usepackage{caption}
\usepackage{subcaption}
\usepackage{enumerate}
\usepackage{multirow}
\newcommand{\algname}[1] {{\fontfamily{cmtt}\selectfont {#1}}}
\definecolor{darkgreen}{rgb}{0.0, 0.5, 0.0}
\usepackage{dsfont}

\usepackage{mathtools}
\DeclarePairedDelimiter{\ceil}{\lceil}{\rceil}
\usepackage[english]{babel}
\usepackage[utf8]{inputenc}
\usepackage{amsmath}
\usepackage{amsfonts}
\usepackage{graphicx}
\usepackage[colorinlistoftodos]{todonotes}
\usepackage{algorithm}
\usepackage{algpseudocode}
\usepackage{siunitx}
\usepackage{tikz}
\usepackage{xcolor}

\begin{document}

\fancyhead{}
  % do not delete this code.

% The "title" command has an optional parameter, allowing the author to define a "short title" to be used in page headers.
\title{FairMatch: A Graph-based Approach for Improving Aggregate Diversity in Recommender Systems}

% The "author" command and its associated commands are used to define the authors and their affiliations.
% Of note is the shared affiliation of the first two authors, and the "authornote" and "authornotemark" commands
% used to denote shared contribution to the research.
\author{Masoud Mansoury}
\authornote{This author also has affiliation in School of Computing, DePaul University, Chicago,
USA, mmansou4@depaul.edu.}
\affiliation{%
  \institution{Eindhoven University of Technology}
  %\streetaddress{P.O. Box 1212}
  \city{Eindhoven}
  \state{the Netherlands}
}
\email{m.mansoury@tue.nl}

\author{Himan Abdollahpouri}
\affiliation{
    \institution{University of Colorado Boulder}
    \city{Boulder}
    \state{USA}
}
\email{himan.abdollahpouri@colorado.edu}

\author{Mykola Pechenizkiy}
\affiliation{%
 \institution{Eindhoven University of Technology}
 \city{Eindhoven}
 \country{the Netherlands}}
\email{m.pechenizkiy@tue.nl}

\author{Bamshad Mobasher}
\affiliation{%
  \institution{DePaul University}
  %\streetaddress{1 Th{\o}rv{\"a}ld Circle}
  \city{Chicago}
  \country{USA}}
\email{mobasher@cs.depaul.edu}

\author{Robin Burke}
\affiliation{%
  \institution{University of Colorado Boulder}
  \city{Boulder}
  \country{USA}}
\email{robin.burke@colorado.edu}

%
% By default, the full list of authors will be used in the page headers. Often, this list is too long, and will overlap
% other information printed in the page headers. This command allows the author to define a more concise list
% of authors' names for this purpose.
\renewcommand{\shortauthors}{Masoud Mansoury, et al.}

%
% The abstract is a short summary of the work to be presented in the article.
\begin{abstract}
Recommender systems are often biased toward popular items. In other words, few items are frequently recommended while the majority of items do not get proportionate attention. That leads to low coverage of items in recommendation lists across users (i.e. low aggregate diversity) and unfair distribution of recommended items. In this paper, we introduce FairMatch, a general graph-based algorithm that works as a post-processing approach after recommendation generation for improving aggregate diversity. The algorithm iteratively finds items that are rarely recommended yet are high-quality and add them to the users' final recommendation lists. This is done by solving the maximum flow problem on the recommendation bipartite graph. While we focus on aggregate diversity and fair distribution of recommended items, the algorithm can be adapted to other recommendation scenarios using different underlying definitions of fairness. A comprehensive set of experiments on two datasets and comparison with state-of-the-art baselines show that FairMatch, while significantly improving aggregate diversity, provides comparable recommendation accuracy.
\end{abstract}

\ccsdesc[500]{Information systems~Personalization}
\ccsdesc[500]{Information systems~Information retrieval diversity}
\ccsdesc[500]{Information systems~Recommender systems}

%
% Keywords. The author(s) should pick words that accurately describe the work being
% presented. Separate the keywords with commas.
\keywords{Recommender Systems, Fairness, Popularity bias, Recommendation coverage, Long-tail, Aggregate diversity}

%
% A "teaser" image appears between the author and affiliation information and the body 
% of the document, and typically spans the page. 
%  \caption{Seattle Mariners at Spring Training, %2010.}
%  \Description{Enjoying the baseball game from the %third-base seats. Ichiro Suzuki preparing to bat.}

%
% This command processes the author and affiliation and title information and builds
% the first part of the formatted document.
\maketitle

\section{Introduction}

Recommender systems are used in a variety of different applications including movies, music, e-commerce, online dating, and many other areas where the number of options from which the user needs to choose can be overwhelming. There are many different metrics to evaluate the performance of the recommender systems ranging from accuracy metrics such as precision, normalized discounted cumulative gain (NDCG), and recall to non-accuracy ones like novelty and serendipity \cite{kaminskas2016}. One of the measures often used to evaluate the effectiveness of a given recommender system is how diverse the list of recommendations given to each user is (aka individual list diversity) \cite{hurley2011}. Recommending a diverse list of items is shown to improve user satisfaction as they give a wider range of options to the user \cite{brynjolfsson2003}. 

The problem with individual list diversity is that it does not capture the extent to which an algorithm covers a diverse set of items across all users which is an important consideration for many applications. Aggregate diversity \cite{adomavicius2011} is a notion to measure this characteristic of the recommender systems and several algorithms have been proposed for that matter by other researchers \cite{adomavicius2011, antikacioglu2017}. Note that a high \textit{individual list diversity} of recommendations does not necessarily imply \textit{high aggregate diversity}. For instance, if the system recommends to all users the same 10 items that are not similar to each other, the recommendation list for each user is diverse (i.e., high individual list diversity), but only 10 distinct items are recommended to all users (i.e., resulting in low aggregate diversity). 

An algorithm with low aggregate diversity could be problematic for several reasons. On the one hand, it concentrates on a limited number of popular items which, in the long run, might negatively affect users' experience in terms of item discovery. Users already know about popular items and recommending them would not add any new information. On the other hand, often items belong to different suppliers and, hence, covering fewer distinct items can indirectly result in an unfair distribution of items across recommendations from the suppliers' perspective. Thus, a low aggregate diversity in recommendation results would have a negative impact on business success and profit \cite{brynjolfsson2011,goldstein2006}.

In this paper, we introduce \textit{FairMatch}, a general graph-based algorithm that works as a post-processing approach after recommendation generation (on top of any existing standard recommendation algorithm) for improving the aggregate diversity. The idea is to generate a list of recommendations with a size larger than what we ultimately want for the final list using a standard recommendation algorithm and then use our FairMatch algorithm to build the final list using a subset of items in the original list. In FairMatch, the main goal is to improve the visibility of high-quality items that have a low visibility in the original set of recommendations. This is done by iteratively solving \textit{Maximum Flow} problem on a recommendation bipartite graph which is built using the recommendations in the original list (left nodes are recommended items and right nodes are the users). At each iteration, the items that can be good candidates for the final list will be selected and removed from the graph, and the process will continue on the remaining part of the graph.

%We explore a special case of supplier fairness under assumption that each item belongs to one supplier and the goal is generating recommendation lists with fair distribution on recommended items. Although in the present paper we concentrate on aggregate diversity and a special case of supplier fairness, our FairMatch algorithm is general and can be applied to recommendation results generated by any standard recommendation algorithm under any definition of fairness. %We will discuss the generality of FairMatch algorithm in the following sections. 

To show the effectiveness of our FairMatch algorithm on improving aggregate diversity and fair visibility of recommended items, we perform a comprehensive set of experiments on recommendation lists of different sizes generated by two standard recommendation algorithms on two publicly available datasets. We intentionally picked two algorithms from two different classes of algorithms (factorization and neighborhood-based models), so our approach is not dependent on any certain type of recommendation algorithms.

Comparison with several state-of-the-art baselines shows that our FairMatch algorithm is able to significantly improve the performance of recommendation results in terms of aggregate diversity and long-tail visibility,
with a negligible loss in the recommendation accuracy in some cases. 

%We provide a comprehensive analysis on this case on how recommendation bipartite graph should be prepared and how weights should be computed. With respect to the power of FairMatch, on multiple recommendation algorithms, datasets, and comparison with two baselines, xQuAD \cite{himan2019a} and FA*IR \cite{zehlike2017}, we observe that FairMatch outperforms xQuAD by ??\% and FA*IR by ??\% in terms of item coverage even with higher recommendation accuracy. 

%We also discuss how FairMatch can be applied for improving other types of fairness on recommendation bipartite graph including individual fairness, group fairness, and other types of provider fairness.  

%\todo[inline]{RB: Biggest addition I would recommend: implement and evaluate a competing baseline algorithm. Otherwise, you aren't showing that you are improving over the state of the art. I would suggest the FA*IR algorithm: https://arxiv.org/pdf/1706.06368. }

\section{Related Work}

The concept of aggregate diversity has been studied by many researchers often under different names such as long-tail recommendation \cite{recsys2017,yin2012challenging}, Matthew effect \cite{moller2018not} and, of course, aggregate diversity \cite{liu2015trust,adomavicius2011improving} all of which refer to the fact that the recommender system should recommend a wider variety of items across all users. 

Vargas and Castells in \cite{vargas2011} proposed probabilistic models for improving novelty and diversity of recommendations by taking into account both relevance and novelty of target items when generating recommendation lists. 
In another work \cite{vargas2014}, they proposed the idea of recommending users to items for improving novelty and aggregate diversity. They applied this idea to nearest neighbor models as an inverted neighbor and a factorization model as a probabilistic reformulation that isolates the popularity components. 

Adomavicius and Kwon \cite{adomavicius2011} proposed the idea of diversity maximization using a maximum flow approach. They used a specific setting for the bipartite recommendation graph in a way that the maximum amount of flow that can be sent from a source node to a sink node would be equal to the maximum aggregate diversity for those recommendation lists. In their setting, given the number of users is $m$, the source node can send a flow of up to $m$ to the left nodes, left nodes can send a flow of up to 1 to the right nodes, and right nodes can send a flow of up to 1 to the sink node. Since the capacity of left nodes to right nodes is set to 1, thus the maximum possible amount of flow through that recommendation bipartite graph would be equivalent to the maximum aggregate diversity.% Although this approach showed improvement on aggregate diversity, there is no control on accuracy of the recommendations and it is possible that increasing aggregate diversity significantly reduces the accuracy of recommendations and there is no way to counterbalance that. 

A more recent graph-based approach for improving aggregate diversity was proposed by Antikacioglu and Ravi in \cite{antikacioglu2017}. They generalized the idea proposed in \cite{adomavicius2011} and showed that the minimum-cost network flow method can be efficiently used for finding recommendation subgraphs that optimizes the diversity. In this work, an integer-valued constraint and an objective function are introduced for discrepancy minimization. The constraint defines the maximum number of times that each item should appear in the recommendation lists and the objective function aims to find an optimal subgraph that gives the minimum discrepancy from the constraint. This work shows improvement in aggregate diversity with a smaller accuracy loss compared to the work in \cite{vargas2011} and \cite{vargas2014}. Similar to this work, our FairMatch algorithm also uses a graph-based approach to improve aggregate diversity. However, unlike the work in \cite{antikacioglu2017} which tries to minimize the discrepancy between the distribution of the recommended items and a target distribution, our FairMatch algorithm has more freedom in promoting high-quality items with low visibility since it does not assume any target distribution of the recommendation frequency.

\section{FairMatch Algorithm}

We formulate our FairMatch algorithm as a post-processing step after the recommendation generation. In other words, we first generate recommendation lists of larger size than what we ultimately desire for each user using any standard recommendation algorithm and use them to build the final recommendation lists. FairMatch works as a batch process, similar to that proposed in \cite{surer2018} where all the recommendation lists are produced at once and re-ranked simultaneously to achieve the objective. In this formulation, we produce a longer recommendation list of size $t$ for each user and then, after identifying high-quality items (items closer to the top of the list) with low visibility (i.e. are not recommended frequently) by iteratively solving the maximum flow problem on recommendation bipartite graph, 
we generate a shorter recommendation list of size $n$ (where $t>>n$). %by applying our FairMatch algorithm to 
%equalize the visibility of recommended items while maintaining the accuracy of recommendations.  
%to provide a more uniform visibility across recommended items while maintaining the accuracy of recommendations.

Let $G=(I,U,E)$ be a bipartite graph of recommendation lists where $I$ is the set of left nodes (representing items), $U$ is the set of right nodes (representing users), and $E$ is the set of edges between left and right nodes showing an item in the left nodes is recommended to a user in the right nodes in recommendation lists of size $t$.  
%In FairMatch, we let $I$ be the set of all items in the original recommendation lists of size $t$, and let $U$ be all users. 
$G$ is initially a uniformly weighted graph, %\todo[]{RB: I would say "uniformly weighted"}
 but we will update the weights for edges as part of our algorithm. %Note that weights play an important role on effectiveness of our algorithm for improving the performance of recommender systems \cite{what do you mean by performance?}. 
We will discuss the initialization and our weighting method in section \ref{weight_comp}. 

Given a weighted bipartite graph $G$, the goal of our FairMatch algorithm is to find high-quality items with low visibility and maximizing their visibility as much as possible without a significant loss in accuracy of the recommendations. Visibility is characterized by the degree of the node in the recommendation graph, while accuracy is captured by the rank position of the items in the original recommendation list. We develop our algorithm by extending the approach introduced in \cite{bonchi2018} to improve the aggregate diversity of the recommender systems. 

We use an iterative process to identify the subgraphs of $G$ that contain the highest quality items with low visibility for each user. After identifying a subgraph $\Gamma$ at each iteration, we remove $\Gamma$ from $G$ and continue the process of finding subgraphs on the rest of the graph (i.e., $G / \Gamma$). We keep track of all the subgraphs as we use them to generate the final recommendations in the last step. 

Identifying $\Gamma$ at each iteration is done by solving a \textit{Maximum Flow} problem (explained in section \ref{fair_max}) on the graph obtained from the previous iteration. Solving the maximum flow problem returns the left nodes connected to the edges with lower weight (i.e., more relevant items with low visibility) on the graph. After finding those left nodes, we form subgraph $\Gamma$ by separating identified left nodes and their connected right nodes from $G$. Finally, $<user,item>$ pairs in subgraphs are used to construct the final recommendation lists of size $n$. We will discuss this process in detail in the following sections.

%\todo[inline]{what do you mean by the last sentence}. 
%Note that while we focus on aggregate diversity and fair item visibility in this paper, FairMatch can be generalized to other definition of fairness. 

%We use a divide-and-conquer strategy to decompose the graph into several parts. At each step that we decompose the graph, the newly decomposed part will contain nodes with the minimum fairness (i.e., items with minimum visibility) that need to be maximized while maintaining overall accuracy \todo[inline]{it is not clear what you mean}. Decomposition \todo[inline]{ we need to be clear in what we mean by decomposition} of the graph at each iteration is done by solving a \textit{Maximum Flow} problem. Based on our definition for fairness (i.e., equalizing the visibility of recommended items) and our weighting strategy for edges in graph, solving the maximum flow problem decomposes the graph \todo[inline]{what do you mean by the last sentence}. Note that while we focus on aggregate diversity and fair item visibility in this paper, FairMatch can be generalized to other definition of fairness \todo[inline]{like how? we can't just claim something with no further explanation. Even if you explain it later in the paper, we still need to say something here }. 

%\todo[inline]{RB: You should make it very clear earlier in the paper that your algorithm works on a batch basis, making all recommendation assignments at once. Maybe even in the abstract.}

%\todo[inline]{RB: Somewhere in this discussion, you should talk about the time complexity of this algorithm.}

Algorithm 1 shows the pseudocode for FairMatch. Overall, our FairMatch algorithm consists of the following four steps: 1) Graph preparation, 2) Weight computation, 3) Candidate selection, and 4) Recommendation list construction. 

\begin{algorithm}
\caption{The FairMatch Algorithm}
\small
\begin{algorithmic}
  \Function{FairMatch}{Recommendations $R$, TopN $n$, Coefficient $\alpha$}
    \State Build graph $G=(I,U,E)$ from $R$
    %\State Assign weight to edges between $L$ and $R$ based on $\mathcal{W}$
    \State Initialize \textit{subgraphs} to empty
    \Repeat
        \State $G$=WeightComputation($G$, $R$, $\alpha$)
        \State $I_{C}$ = Push-relabel($G$)
        \State Initialize $subgraph$ to empty
        \For {\textbf{each} $i \in I_{C}$}
            \If{$label_{i} \geq |I|+|U|+2$}
                \For {\textbf{each} $u \in Neighbors(i)$}
                    \State Append $<i,u,e_{iu}>$ to \textit{subgraph}
                \EndFor
            \EndIf 
        \EndFor
        \If{$subgraph$ is empty}
            \State $break$
        \EndIf
        \State Append $subgraph$ to $subgraphs$
        \State $G$=Remove $subgraph$ from $G$
    \Until($true$)
    \State Reconstruct R of size $n$ based on \textit{subgraphs}
   \EndFunction
\end{algorithmic}
%\begin{tikzpicture}[yscale=0.5]
%    \draw [line width=0mm,   black] (0,-5) -- (8.1,-5) node [right];
%\end{tikzpicture}
%\begin{algorithmic}
%    \Function{WeightComputation}{Graph $G=(I,U,E)$, Recommendation $R$, %Trade-off coefficient $\alpha$}
%        \For {\textbf{all} edges $e_{i\in I,u\in U} \in E$}
%            \State Set $relevance(i,u)$ to the position of item $i$ in %$R_{u}$
%            \State $w_{iu}=\alpha \times degree(i) + (1-\alpha) \times %relevance(i,u)$
%        \EndFor
%        \For {\textbf{all} edges $e_{s,i\in I} \in E$}
%            \State Compute $w_{si}$
%        \EndFor
%        \For {\textbf{all} edges $e_{u\in U,t} \in E$}
%            \State Compute $w_{ut}$
%        \EndFor
%        \State \textbf{return} updated $G$
%    \EndFunction
%\end{algorithmic}
\end{algorithm}

\begin{figure}[h]
    \centering
    \includegraphics[width=0.25\textwidth]{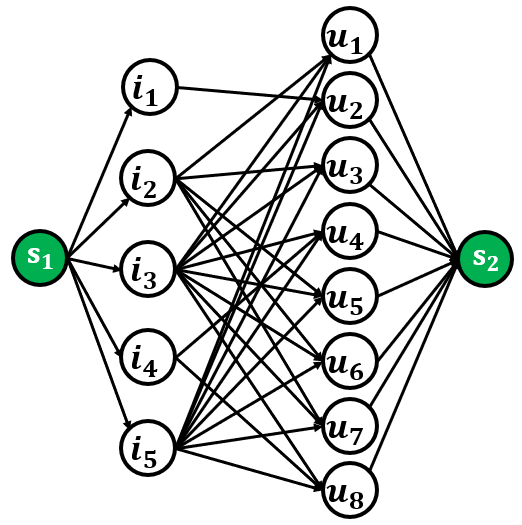}
    \caption{An example of a recommendation bipartite graph of recommendation lists of size 3.}\label{fig:ex}
\end{figure}

\subsection{Graph Preparation}

Given long recommendation lists of size $t$ generated by a standard recommendation algorithm, we create a bipartite graph from recommendation lists in which items and users are the nodes (called, respectively, left and right nodes) and recommendations are expressed as edges. Since our FairMatch algorithm is formulated as a maximum flow problem, we also add two nodes, \textit{source} ($s_1$) and \textit{sink} ($s_2$). The purpose of having a source and sink node in the maximum flow problem is to have a start and endpoint for the flow going through the graph. We connect $s_1$ node to all left nodes and also we connect all right nodes to $s_2$. Figure \ref{fig:ex} shows a sample bipartite graph resulted in this step. 

%Assuming that the goal of the recommendation algorithm is generating a list of size $n$ preferred items for each user, first we generate a longer recommendation list of size $t >> n$ preferred items for each user. We create a bipartite graph from this long recommendation lists where the left nodes are the recommended items, right nodes are the users, and edges between an item and a user indicate the item is recommended to the user. 

%In addition, we add two special nodes, \textit{source node} $s$ and \textit{sink node} $t$, which are necessary for solving maximum flow problem on the graph \todo[inline]{why is it necessary?}. We connect $s$ to all the left nodes. Similarly, we connect all the right nodes to $t$.

%\input{t_weighting.tex}

\subsection{Weight Computation}\label{weight_comp}

Given the bipartite recommendation graph, $G=(I,U,E)$, the task of weight computation is to calculate the weight for edges between the source node and left nodes, left nodes and right nodes, and right nodes and sink node.

For edges between left nodes and right nodes, we define the weights as the weighted sum of item visibility and relevance. The visibility of each item is defined as the degree of the node corresponding to that item (excluding the edge with the source node). Item degree is the number of edges going out from that node connecting it to the user nodes and that shows how often it is recommended to different users. Relevance is based on the rank of the item in the original recommendation list for each user (lower rank is more relevant).

For computing the weight between $i \in I$ and $u \in U$, we use the following equation:

\begin{equation}\label{w}
    w_{iu}= \alpha \times degree_{i} + (1 - \alpha) \times rank_{iu}
\end{equation}

\noindent where $degree_{i}$ is the number of edges from $i$ to right nodes (i.e., $u \in U$), $rank_{iu}$ is the position of item $i$ in the recommendation list of size $t$ generated for user $u$, and $\alpha$ is a coefficient to control the trade-off between accuracy and diversity (or visibility).  

\begin{figure*}[htp]
    \centering
    \begin{subfigure}[b]{0.26\textwidth}
        \includegraphics[width=\textwidth]{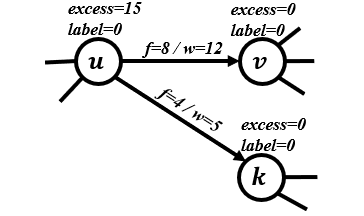}
        \caption{Original graph} \label{fig:push_ex:a}
    \end{subfigure}
    \begin{subfigure}[b]{0.27\textwidth}
        \includegraphics[width=\textwidth]{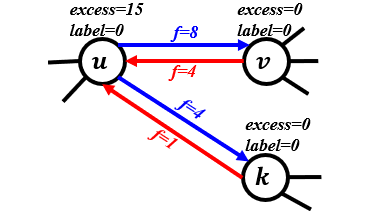}
        \caption{Residual graph} \label{fig:push_ex:b}
    \end{subfigure}%
    \begin{subfigure}[b]{0.27\textwidth}
        \includegraphics[width=\textwidth]{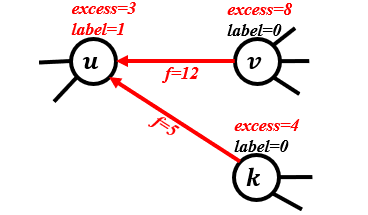}
        \caption{Pushing excess flow of u} \label{fig:push_ex:c}
    \end{subfigure}%
\caption{Example of push and relabel operation.} \label{fig:push_ex}
\end{figure*}

Note that in equation \ref{w}, $degree$ and $rank$ have different ranges. The range for $rank$ is from 1 to $t$ (there are $t$ different positions in the original list) and the range of $degree$ depends on the frequency of the item recommended to the users (the more frequent it is recommended to different users the higher its degree is). Hence, for a meaningful weighted sum, we normalize $degree$ to be in the same range as $rank$. 
%where $c$ is a coefficient to control the magnitude of weights and its value will experimentally be tuned. $\mathcal{W}(e)$ returns the weight for edge $e$ connected to $l$ based on specified weighting strategy.  

Given weights of the edges between $i \in I$ and $u \in U$, $w_{iu}$, total capacity of $I$ and $U$ would be %as follow:  
$C_{T}=\sum_{i\in I}^{}\sum_{u \in U}^{}w_{iu}$
%\begin{equation} \label{eq:sum_cap}
%C_{T}=\sum_{i\in I}^{}\sum_{u \in U}^{}w_{iu}
%\end{equation}
which simply shows the sum of the weights of the edges connecting left nodes to the right nodes.

For computing the weight for edges connected to the source and sink nodes, first, we equally distribute $C_{T}$ to left and right nodes. Therefore, the capacity of each left node, $C_{eq}(I)$, and right node, $C_{eq}(U)$, would be as follow: 

\begin{equation}
C_{eq}(I)=\ceil[\bigg]{\frac{C_{T}}{|I|}}, \;\;\;\;\;\; C_{eq}(U)=\ceil[\bigg]{\frac{C_{T}}{|U|}}
\end{equation}
%\begin{equation}
%C_{eq}(U)=\ceil[\bigg]{\frac{C_{T}}{|U|}} 
%\end{equation}

\noindent where $\ceil[\big]{a}$ returns the ceil value of $a$.
%For example, suppose the total capacity, $C_{T}$, is 100. If we have 5 left nodes and 8 right nodes (similar to Figure \ref{fig:ex}), then the capacity of each left node would be 20 ($\ceil[\big]{100/5}$) and the capacity of each right node would be 13 ($\ceil[\big]{100/8}$).
%For example, in figure \ref{fig:ex}, using equation \ref{eq:sum_cap} and based on the weights computed for edges between left and right nodes, $C_T$ is equal to 100. Then, 
Then, based on equal capacity assigned to each left and right nodes, we follow the method introduced in \cite{bonchi2018} to compute weights for edges connected to source and sink nodes as follow:
%For computing the weight for edges connected to source and sink nodes, we follow the method introduced in \cite{bonchi2018}. For these edges, we equally distribute the total capacity of the graph as follow:

\begin{equation} \label{eq:sl_cap1}
\forall i \in I, w_{s_{1}i}=\ceil[\bigg]{min(\frac{C_{eq}(I)}{gcd(C_{eq}(I),C_{eq}(U))},\frac{C_{eq}(U)}{gcd(C_{eq}(I),C_{eq}(U))})}
\end{equation}

%\begin{equation}
%    \ceil[\big]{x} \ceil[\Big]{x} \ceil[\bigg]{x} \ceil[\Bigg]{x}
%\end{equation}

\begin{equation} \label{eq:sl_cap2}
\forall u \in U, w_{us_{2}}=\ceil[\bigg]{\frac{C_{eq}(I)}{gcd(C_{eq}(I),C_{eq}(U))}}
\end{equation}

\noindent where $gcd(C_{eq}(I),C_{eq}(U))$ is the Greatest Common Divisor of the distributed capacity of left and right nodes. %(e.g. $gdc(13,25)=1$ in our previous example). 
%In this computation, $gcd(C_{eq}(I),C_{eq}(U))$ signifies the highest possible amount of capacity that equally can be passed through each node. Also, 
Assigning the same weight to edges connected to the source and sink nodes guaranties that all nodes in $I$ and $U$ are treated equally and the weights between them play an important role in our FairMatch algorithm. 

\subsection{Candidate Selection}\label{fair_max}

The graph constructed in previous steps is ready to be used for solving the maximum flow problem. In a maximum flow problem, the main goal is to find the maximum amount of feasible flow that can be sent from the source node to the sink node through the flow network. Several algorithms have been proposed for solving a maximum flow problem. Well-known algorithms are Ford--Fulkerson \cite{ford1956}, Push-relabel \cite{goldberg1988}, and Dinic's algorithm \cite{dinic1970}. In this paper, we use Push-relabel algorithm to solve the maximum flow problem on our bipartite recommendation graph as it is one of the efficient algorithms for this matter.

In push-relabel algorithm, each node will be assigned two attributes: \textit{label} and \textit{excess flow}. The label attribute is an integer value that is used to identify the neighbors to which the current node can send flow. A node can only send flow to neighbors that have lower label than the current node. Excess flow is the remaining flow of a node that can still be sent to the neighbors. When all nodes of the graph have excess flow equals to zero, the algorithm will terminate.

The push-relabel algorithm combines $push$ operations that send a specific amount of flow to a neighbor, and $relabel$ operations that change the label of a node under a certain condition (when the node has excess flow greater than zero and there is no neighbor with label lower than the label of this node).

Here is how the push-relabel algorithm works: Figure \ref{fig:push_ex} shows a typical graph in the maximum flow problem and an example of push and relabel operations. In Figure \ref{fig:push_ex:a}, $f$ and $w$ are current flow and weight of the given edge, respectively. In Push-relabel algorithm, a residual graph, $G^{'}$, will be also created from graph $G$. As graph $G$ shows the flow of forward edges, graph $G^{'}$ shows the flow of backward edges 
%\todo[inline]{check whether this is true. The figure does not seem like this} 
 calculated as $f_{backward}=w-f$. Figure \ref{fig:push_ex:b} shows residual graph of graph $G$ in Figure \ref{fig:push_ex:a}. Now, we want to perform a push operation on node $u$ and send its excess flow to its neighbors. 

Given $x_{u}$ as excess flow of node $u$, $push(u,v)$ operation will send a flow of amount $\Delta=min(x_{u},f_{uv})$ from node $u$ to node $v$ and then will decrease excess flow of $u$ by $\Delta$ (i.e., $x_{u}=x_{u}-\Delta$) and will increase excess flow of $v$ by $\Delta$ (i.e., $x_{v}=x_{v}+\Delta$). After $push(u,v)$ operation, node $v$ will be put in a queue of active nodes to be considered by the push-relabel algorithm in the next iterations and residual graph would be updated. Figure \ref{fig:push_ex:c} shows the result of $push(u,v)$ and $push(u,k)$ on the graph shown in Figure \ref{fig:push_ex:b}. In $push(u,v)$, for instance, since $u$ and all of its neighbors have the same label value, in order to perform push operation, first we need to perform relabel operation on node $u$ to increase the label of $u$ by one unit more than the minimum label of its neighbors to guaranty that there is at least one neighbor with lower label for performing push operation. After that, node $u$ can send flow to its neighbors. 

Given $x_{u}=15$, $f_{uv}=8$, and $f_{uk}=4$ in Figure \ref{fig:push_ex:b}, after performing relabel operation, we can only send the flow of amount 8 from $u$ to $v$ and the flow of amount 4 from $u$ to $k$. After these operations, residual graph (backward flow from $v$ and $k$ to $u$) will be updated. 

The push-relabel algorithm starts with a "preflow" operation to initialize the variables and then it iteratively performs push or relabel operations until no active node exists for performing operations. Assuming $\mathcal{L}_v$ as the label of node $v$, in preflow step, we initialize all nodes as follow: $\mathcal{L}_{s_1}=|I|+|U|+2$, $\mathcal{L}_{i \in I}=2$, $\mathcal{L}_{u \in U}=1$, and $\mathcal{L}_{s_2}=0$. This way, we will be able to send the flow from $s_1$ to $s_2$ as the left nodes have higher label than the right nodes. Also, we will push the flow of amount $w_{s_{1}i}$ (where $i \in I$) from $s_1$ to all the left nodes. 

After preflow, all of the left nodes $i \in I$ will be in the queue, $\mathcal{Q}$, as active nodes because all those nodes now have positive excess flow. The main part of the algorithm will now start by dequeuing an active node $v$ from $\mathcal{Q}$ and performing either push or relabel operations on $v$ as explained above. This process will continue until $\mathcal{Q}$ is empty. At the end, each node will have specific label value and the sum of all the coming flows to node $s_2$ would be the maximum flow of graph $G$. For more details see \cite{goldberg1988}%\footnote{You can interactively run Push-relabel algorithm on your graph here: http://www.adrian-haarbach.de/idp-graph-algorithms/implementation/maxflow-push-relabel/index\_en.html}. 

An important question is: \textit{how does the Push-relabel algorithm can find high-quality nodes (items) with low degree (visibility)?} 

We answer this question by referring to the example in Figure \ref{fig:push_ex:c}. In this figure, assume that $u$ has a backward edge to $s_1$. Since $u$ has excess flow greater than zero, it should send it to its neighbors. However, as you can see in the figure, $u$ does not have any forward edge to $v$ or $k$ nodes. Therefore, it has to send its excess flow back to $s_1$ as $s_1$ is the only reachable neighbor for $u$. Since $s_1$ has the highest label in our setting, in order for $u$ to push all its excess flow back to $s_1$, it should go through a relabel operation so that its label becomes larger than that of $s_1$. Therefore, the label of $u$ will be set to $\mathcal{L}_{s_1}+1$ for an admissible push.

The reason that $u$ receives high label value is the fact that it initially receives high flow from $s_1$, but it does not have enough capacity (the sum of weights between $u$ and its neighbors is smaller than its excess flow. i.e. 8+4<15) to send all that flow to them. In FairMatch, in step 3 (i.e. section \ref{fair_max}), left nodes without sufficient capacity on their edges will be returned as part of the outputs from push-relabel algorithm and are considered for constructing the final recommendation list in step 4 (i.e. section \ref{step4}). These nodes are the ones that their edges received low weights by equation \ref{w} in step 2 (i.e. section \ref{weight_comp}) because of their low degree (low visibility) and rank (high relevance) on the graph. Therefore, FairMatch aims at promoting those high relevance items with low visibility.

\subsection{Recommendation List Construction}\label{step4}

In this step, the goal is to construct a recommendation list of size $n$ by the $<user,item>$ pairs identified in previous step. %For simplicity, we reconstruct original recommended lists by adding the identified nodes by FairMatch to the end of the lists. 

Given a recommendation list of size $n$ for user $u$, $R_{u}$, sorted based on the scores generated by a standard recommendation algorithm, candidate items identified by FairMatch connected to $u$ as $\mathcal{I}_{C}$, and visibility of each item, $i$, in recommendation lists of size $n$ as $V_{i}$, we use the following process for generating recommendation list for $u$. First, we sort recommended items in $R_{u}$ based on their $V_{i}$ in ascending order. Then, we remove $min(n,|\mathcal{I}_{C}|)$ from the bottom of sorted $R_{u}$ and add $min(n,|\mathcal{I}_{C}|)$ items from $I_{C}$ to the end of $R_{u}$.

This process will ensure that extracted items in the previous step will replace the frequently recommended items meaning that it decreases the visibility of the frequently recommended items and increases the visibility of rarely recommended items to generate a fair distribution on recommended items.

\section{Experiments}

We performed a comprehensive evaluation of the effectiveness of FairMatch in improving aggregate diversity of recommender systems. Our evaluation on two standard recommendation algorithms and comparison to various diversification methods to increase aggregate diversity as baselines on two datasets shows that FairMatch significantly improves item visibility with a negligible loss in the accuracy of recommendations. 

\subsection{Experimental Setup}

Experiments are performed on two publicly available datasets: Epinions and MovieLens. The Epinions dataset  %\footnote{http://www.trustlet.org/downloaded\_epinions.html} dataset \cite{Massa:2007a} 
was collected from Epinions web site which is an item reviewing system. It is a subset extracted from Epinions dataset in which each user has rated at least 15 items and each item is rated by at least 15 users (i.e core-15). The MovieLens dataset \cite{harper2015} is movie ratings data and was collected by the GroupLens %\footnote{https://grouplens.org/datasets/movielens/} 
research group. The characteristics of the datasets are summarized in Table \ref{tab:dataset}. %These datasets are from various domains, have different level of sparsity, and are different in terms of percentage of long-tail items within the dataset.

\captionsetup[table]{skip=4pt}
\begin{table}[t!]
\small
\centering
\captionof{table}{Statistical properties of datasets} \label{tab:dataset}
\begin{tabular}{lrrrr}
    \toprule
     Dataset & \#users & \#items & \#ratings & density \\
     \midrule
     Epinions & 5,531 & 4,287 & 186,995 & 0.789\% \\
     ML1M & 6,040 & 3,706 & 1,000,209 & 4.468\% \\
    \bottomrule
\end{tabular}
\end{table}

%The BX dataset is a subset extracted from the BookCrossing dataset\footnote{http://www.informatik.uni-freiburg.de/~cziegler/BX/} such that each user has rated at least 5 books and each book is rated by at least 5 users. The Epinions dataset\footnote{http://www.trustlet.org/downloaded_epinions.html} contains users' ratings on different products and was collected from Epinions website. In Epinions dataset, each user has rated at least ?? products and each product is rated by at least ?? users. Finally, the ML1M is movie ratings data and was collected by the MovieLens\footnote{https://grouplens.org/datasets/movielens/} research group.  

The initial longer recommendation lists of size $t$ are generated by two well-known recommendation algorithms: list-wise matrix factorization (\algname{ListRank}) \cite{shi2010} and user-based collaborative filtering (\algname{UserKNN}) \cite{Resnick:1994a}. As mentioned earlier, we chose these two algorithms to cover different approaches in recommender systems: matrix factorization and neighborhood models. We performed gridsearch\footnote{For \algname{ListRankMF}, we set all regularizers $\in \{0.0001,0.001,0.01\}$, $iterations \in \{30,50,100\}$, $learning rate \in \{0.0001,0.001,0.005,0.01\}$, and $factors \in \{50,100,150,200\}$. For \algname{UserKNN}, we set $neighbors \in \{10,20,30,50,100\}$.} on hyperprameters for each algorithm and selected the results with the highest precision value for our next analysis.

To show the effectiveness of the FairMatch algorithm in improving the aggregate diversity, we compare its performance with two state-of-the-art algorithms and also two simple baselines. 

\input{t_ep.tex}
\input{t_ml1m.tex}

\begin{enumerate}
  \item \textbf{FA*IR.} This is the method introduced in \cite{zehlike2017} and was mentioned in our related work section. The method was originally used for improving group fairness in job recommendation. However, we use this method for improving aggregate diversity in item recommendation. We define protected and unprotected groups as long-tail and short-head items, respectively. For separating short-head from long-tail items, we consider those top items which cumulatively take up K\% of the ratings as the short-head and the rest as long-tail items. For experiments in this paper, we have tried different values of $K \in \{10,20,30,50\}$. 
  %For specifying long-tail items from training data, we create cumulative popularity list of items sorted from most popular to less-popular items, then we define a cutting point such that it divides the items into short-head and long-tail items with a certain ratio (e.g., cutting point of 20 means that 20\% of the items belong to short head and the remaining 80\% fall in into the long-tail). 
%   %For experiments in this paper, we tested various cutting points as $\{10,20,30,50\}$. 
Also, we set the other two hyperparameters, proportion of protected candidates in the top $n$ items\footnote{Based on suggestion from the released code, the range should be in $[0.02,0.98]$} and significance level\footnote{Based on suggestion from the released code, the range should be in $[0.01,0.15]$}, to $\{0.25,0.5,0.75,0.95\}$ and $\{0.05,0.1,0.15\}$, respectively. 
  
  \item \textbf{Discrepancy Minimization (DM)}. This is the method introduced in \cite{antikacioglu2017} and was explained in our related work section. For hyperparameter tuning, we followed the experimental settings suggested by the original paper for our experiments. We set the target degree distribution to $\{1,5,10\}$ and relative weight of the relevance term to $\{0.01,0.5,1\}$.  
  %\todo[inline]{This table has an AGG algorithm. Should be here also.}
  
  \item \textbf{Reverse.} Given a recommendation list of size $t$ for each user generated by standard recommendation algorithm, in this method, instead of picking the $n$ items from the top, we pick them from the bottom of the list.  In this approach, we expect to see an increase in aggregate diversity as we are giving higher priority to the items with lower scores to be picked first. However, the accuracy of the recommendations will decrease as we give higher priority to lower quality items.
  
  \item \textbf{Random.} Given a recommendation list of size $t$ for each user generated by standard recommendation algorithm, we randomly choose $n$ items from that list and create a final recommendation list for that user. Note that this is different from randomly choosing items from all catalog to recommend to users. The reason we randomly choose the items from the original recommended list of items (size $t$) is to compare other post-processing and re-ranking techniques with a simple random re-ranking. 
\end{enumerate}

FairMatch algorithm only involves one hyperparameter, $\alpha$, to control the balance between the node degree and relevance. For our experiments we try $\alpha \in \{0,0.25,0.5,0.75,1\}$. A lower value for $\alpha$ indicates more focus on maintaining the accuracy of the recommendations, while a higher value for $\alpha$ indicates more focus on improving aggregate diversity. We also perform a sensitivity analysis to show how $\alpha$ can play an important role in the accuracy-diversity trade-off.

For evaluation, we use the following metrics to measure the effectiveness of each method:

\begin{enumerate}
    \item \textbf{Precision ($P@n$)}: The fraction of the recommended items shown to the users that are part of the users' profile in the test set.
    \item \textbf{Coverage ($C@n$)}: The percentage of items which appear at least once in the recommendation lists.
    \item \textbf{Gini index ($G@n$)}: The measure of fair distribution of recommended items. It takes into account how uniformly items appear in recommendation lists. Uniform distribution will have Gini index equal to zero which is the ideal case (lower Gini index is better). Given all the recommendation lists for users, $L$, and $p(i_{k}|L)$ as the probability of the $k$-th least recommended item being drawn from $L$ calculated as \cite{vargas2014}:
    
    \begin{equation}
        p(i|L)=\frac{\sum_{u \in U}^{} \mathds{1}_{i \in L_{u}}}{\sum_{u \in U}^{} \sum_{j \in I}^{} \mathds{1}_{j \in L_{u}}} 
    \end{equation}
    
    where $L_{u}$ is the recommendation list for user $u$. Now, Gini index of $L$ can be computed as:
    
    \begin{equation}
        Gini(L)=\frac{1}{|I|-1} \sum_{k=1}^{|I|} (2k-|I|-1)p(i_{k}|L) 
    \end{equation}
    \item \textbf{Entropy ($E@n$)}: Given the distribution of recommended items, entropy measures the uniformity of that distribution. Uniform distribution has the highest entropy or information gain, thus higher entropy is more desired when the goal is increasing diversity. 
    
    \begin{equation}
        Entropy(L)=- \sum_{i \in I}^{} p(i|L) \log p(i|L) 
    \end{equation}
    
    where $p(i|L)$ is the observed probability value of item $i$ in recommendation lists $L$.
\end{enumerate}

We performed 5-fold cross validation in our experiments, and we generated recommendation lists of size 10, 20, 50, and 100 for each user by each recommendation algorithm. Recommendation lists of size 10 are used for evaluating standard recommendation algorithms and longer recommendation lists of size 20, 50, and 100 are used as input for diversification techniques to generate recommendation lists of size 10 as output. Recommendation lists of size 10 generated by each diversification technique are evaluated by aforementioned metrics and their effectiveness is compared. We used \textit{librec-auto} and LibRec 2.0 for running experiments \cite{mansoury2018automating,Guo2015}.

\subsection{Comparative Evaluation}

Table \ref{tab:ep} and \ref{tab:ml} summarize the performance of FairMatch and other baselines on Epinions and MovieLens datasets, respectively. For each metric (ignoring Random and Reverse techniques), the bolded values show the best results and a statistically significant change from the second best baseline with $p<0.01$. %Italicized values show a statistically insignificant change from the second best baseline with $p<0.01$.

As mentioned earlier, extensive experiments are performed by each diversification technique with multiple hyperparameter values and for the purpose of comparison, from each of three diversification algorithms (DM, FA*IR, and our FairMatch) the configuration which yields, more of less, the same precision loss is reported. These results enable us to better compare the performance of each technique on improving aggregate diversity while maintaining the same level of accuracy. 

Based on experiments on Epinions dataset shown in table \ref{tab:ep}, FairMatch significantly outperforms all the baselines on various sizes of initial recommendation lists generated by both recommendation algorithms in terms of coverage ($C@10$). The coverage of FairMatch is even higher than the Random and Reverse techniques without losing much accuracy which is indicative of its power in finding high-quality items with minimum visibility. Again, the $Random$ algorithm used here is randomly picking $n$ items from the original list and put them in the final list, so it is still possible that many popular items could end up being in the final list. In terms of fair distribution, the same improvement is also consistently observed on entropy. Entropy of FairMatch technique is significantly higher than other techniques in all cases showing that the recommendations generated by FairMatch are fairer and closer to a uniform distribution. However, in terms of Gini index, FairMatch generated comparable results to DM.    

\begin{figure}[htp]
    \centering
    \begin{subfigure}[b]{0.45\textwidth}
        \includegraphics[width=\textwidth]{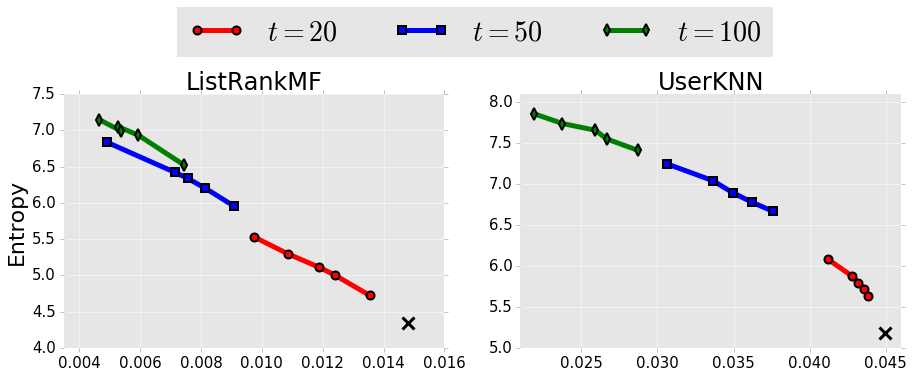}
        \caption{Epinions dataset} \label{fig:sensitivity:a}
    \end{subfigure}
    \begin{subfigure}[b]{0.45\textwidth}
        \includegraphics[width=\textwidth]{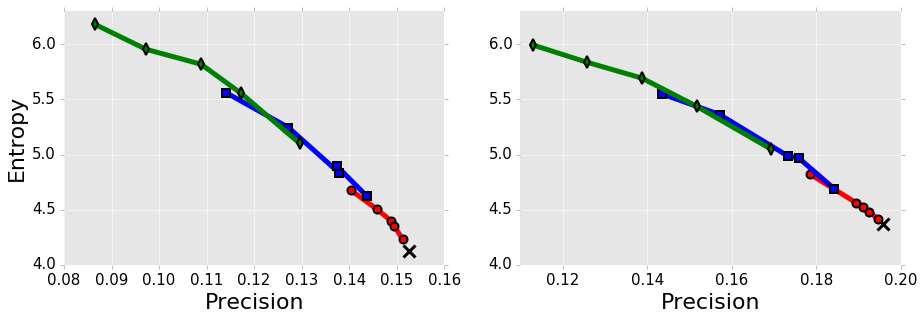}
        \caption{MovieLens dataset} \label{fig:sensitivity:b}
    \end{subfigure}%
\caption{Precision and entropy trade-off in the FairMatch algorithm on Epinions and MovieLens datasets using \algname{ListRankMF} and \algname{UserKNN}. The black cross shows the performance of original recommendation lists at size 10.} \label{fig:sensitivity}
\end{figure}

Table \ref{tab:ml} shows the experimental results in MovieLens dataset. Based on these results, except for \algname{UserKNN} with $t=50$ and $t=100$, FairMatch provides higher coverage in all cases which is consistent with the results from Epinions dataset. In terms of entropy and Gini index, FairMatch was outperformed by DM in most of the cases.

%To further analyze the results obtained from Epinions and MovieLens dataset, it is clear that FairMatch provided better results on Epinions dataset in terms of fair distribution on recommended items. One possible reason can be the distribution of recommended items in long recommendation lists. Image ?? shows the distribution of recommended items on both datasets for recommendation lists of size 20, 50, and 100. As shown, the situation on Epinions dataset is more severe as majority of items do not receive sufficient visibility, while few items are frequently appeared in recommendation lists. On the other hand, on MovieLens dataset, the distribution is more 

It is worth noting that the Gini can be a misleading measure if it is looked at in isolation. For instance, if an algorithm recommends only a few items (low coverage) but does so by recommending each item exactly in an equal proportion, then it will achieve a perfect Gini. However, having a low coverage is not desired and therefore it is more reasonable to look at the coverage and Gini together.  

\subsection{Accuracy-Diversity Trade-Off}

We also investigated the precision and diversity trade-off in our FairMatch algorithm under various settings. Figure \ref{fig:sensitivity} shows the experimental results on Epinions (Figure \ref{fig:sensitivity:a}) and MovieLens (Figure \ref{fig:sensitivity:b}) datasets. In these plots, x-axis shows the precision and y-axis shows the entropy of the recommendation results at size 10. Similar results are also observed when Gini index or coverage metrics are used as diversity measures. Each point on the plot corresponds to a specific $\alpha$ value and the black cross shows the performance of original recommendation lists at size 10.

Results in Figure \ref{fig:sensitivity} show that $\alpha$ plays an important role in controlling the precision-diversity trade-off. As we increase the $\alpha$ value, precision increases, while diversity decreases. According to equation \ref{w}, for a higher $\alpha$ value, FairMatch will concentrate more on improving the accuracy of the recommendations, while for lower $\alpha$ value, it will have a higher concentration on improving the diversity of the recommendations.

Also, it can be observed from Figure \ref{fig:sensitivity} that for longer initial recommendation lists (i.e., higher values for $t$), although the diversity of the recommendations increases, the precision decreases. These parameters allow system designers to better control the precision-diversity trade-off.

%\begin{figure}[htp]
%    \centering
%    \begin{subfigure}[b]{0.34\textwidth}
%        \includegraphics[width=\textwidth]{lt-ep-listrank.png}
%    \end{subfigure}
%    \begin{subfigure}[b]{0.34\textwidth}
%        \includegraphics[width=\textwidth]{lt-ep-userknn.png}
%    \end{subfigure}%
%\caption{Epinions dataset.} \label{fig:lt-ep}
%\end{figure}
%\begin{figure}[htp]
%    \centering
%    \begin{subfigure}[b]{0.34\textwidth}
%        \includegraphics[width=\textwidth]{lt-ml-listrank.png}
%    \end{subfigure}
%    \begin{subfigure}[b]{0.34\textwidth}
%        \includegraphics[width=\textwidth]{lt-ml-userknn.png}
%    \end{subfigure}
%\caption{MovieLens dataset.} \label{fig:lt-movie}
%\end{figure}

%\begin{figure}[htp]
%    \centering
%    \begin{subfigure}[b]{0.5\textwidth}
%        \includegraphics[width=0.5\textwidth]{lt-ep-listrank.png}
%        \includegraphics[width=0.5\textwidth]{lt-ep-userknn.png}
%        \caption{Epinions dataset} \label{fig:lt-ep}
%    \end{subfigure}
%    \begin{subfigure}[b]{0.5\textwidth}
%        \includegraphics[width=0.5\textwidth]{lt-ml-listrank.png}
%        \includegraphics[width=0.5\textwidth]{lt-ml-userknn.png}
%        \caption{MovieLens dataset} \label{fig:lt-ep}
%    \end{subfigure}
%\caption{Caption} \label{fig:lt}
%\end{figure}

\begin{figure}[htp]
    \centering
    \begin{subfigure}[b]{0.45\textwidth}
        \includegraphics[width=\textwidth]{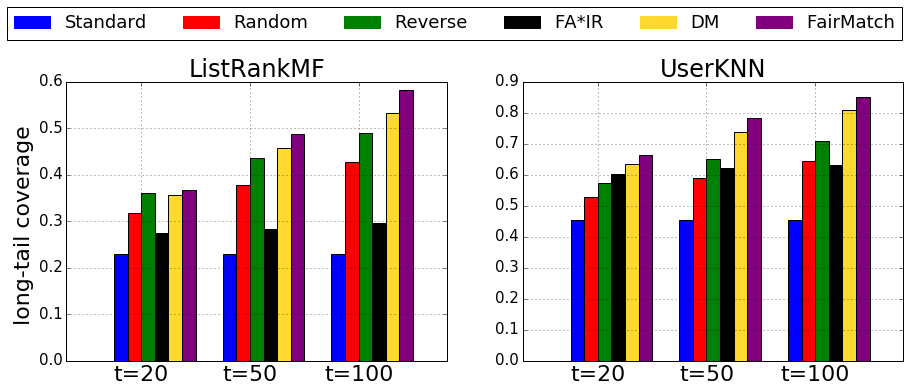}
        \caption{Epinions dataset} \label{fig:lt:a}
    \end{subfigure}
    \begin{subfigure}[b]{0.45\textwidth}
        \includegraphics[width=\textwidth]{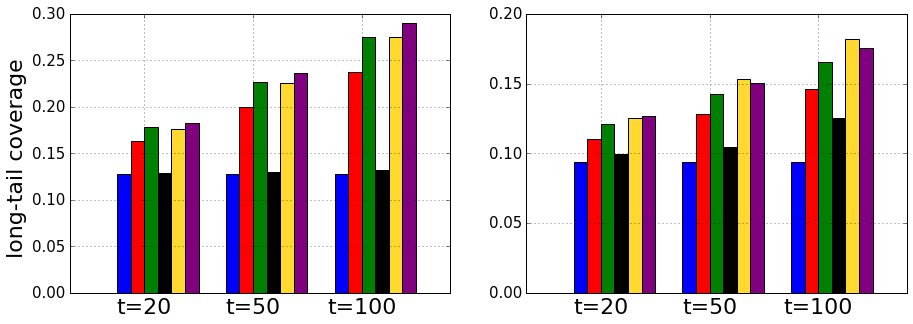}
        \caption{MovieLens dataset} \label{fig:lt:b}
    \end{subfigure}%
\caption{Long-tail coverage of diversification methods on Epinions and MovieLens datasets using \algname{ListRankMF} and \algname{UserKNN}. } \label{fig:lt}
\end{figure}

%\begin{figure*}[ht]
%    \centering
%    \includegraphics[width=\textwidth]{ep-sensitivity.jpg}
%    \caption{Sensitivity analysis.} \label{fig:push_ex}
%\end{figure*}
%\begin{figure*}[ht]
%    \centering
%    \includegraphics[width=\textwidth]{ml-sensitivity.jpg}
%\caption{Sensitivity analysis.} \label{fig:sensitivity}
%\end{figure*}
\subsection{Long-tail Coverage Analysis}

Recommending more items by a given recommendation algorithm is a desired characteristic. However, it is important to check if the increase in item coverage comes from recommending more long-tail items or it is just covering more popular items. Figure ~\ref{fig:lt} shows the long-tail coverage for different algorithms on Epinions (Figure ~\ref{fig:lt:a}) and MovieLens (Figure ~\ref{fig:lt:b}) datasets for different original recommendations sizes $t$. For these experiments, we specified long-tail items using the technique introduced in \cite{celma2008}. Except for the \algname{UserKNN} on MovieLens dataset which our FairMatch algorithm covers fewer long-tail items than the DM algorithm, in all other cases, the FairMatch algorithm outperforms all other algorithms on both datasets. In fact, on MovieLens, the FairMatch algorithm also beats DM algorithm with a slight margin when the size of the original recommendation is 20. In other words, when the time and space complexity become an issue (larger values for $t$) and a smaller $t$ is desired then the FairMatch algorithm outperforms every other algorithm in this experiment on both datasets.

\subsection{Complexity Analysis}

Solving the maximum flow problem is the core computation part of the FairMatch algorithm. We used Push-relabel algorithm as one of the efficient algorithms for solving the maximum flow problem. This algorithm has a polynomial time complexity as $O(V^{2}E)$ where $V$ is the number of nodes and $E$ is the number of edges in bipartite graph. For other parts of the FairMatch algorithm, the time complexity would be in the order of the number of edges as it mainly iterates over the edges in the bipartite graph.  

Since FairMatch is an iterative process, unlike other maximum flow based techniques \cite{adomavicius2011, antikacioglu2017}, it requires solving maximum flow problem on the graph multiple times and this could be one limitation of our work. However, except for the first iteration that FairMatch executes on the original graph, at the next iterations, the graph will be shrunk as FairMatch removes some parts of the graph at each iteration. Regardless, the upper-bound for the complexity of FairMatch will be $O(V^{3}E)$ assuming in each iteration we still have the entire graph (which is not the case). Therefore, the complexity of FairMatch is certainly less than $O(V^{3}E)$ which is still polynomial. 
%One limitation of FairMatch in terms of time complexity is its iterative
%In comparison to DM as one of the competitive and most relevant baselines to FairMatch which also uses flow network for improving aggregate diversity, FairMatch is computationally more expensive than DM.   

\section{Discussion and Future Work}

In this section, we discuss the advantages that FairMatch provides on improving the performance of recommender systems. Also, we will discuss possible future work that can be considered for further improvement in FairMatch.

\textbf{Generalization.} In this paper, we studied the ability of FairMatch for improving aggregate diversity and one special case of supplier fairness under the assumption of each item belongs to one supplier (i.e., fair distribution on recommended items). However, FairMatch can be generalized to other definitions of fairness including supplier-side fairness. In this scenario, we can create recommendation bipartite graph between users and suppliers (based on recommended items), and then assign weights to edges based on suppliers' information (e.g., the probability of their items being shown in recommendation results and the quality of their items). At the third step, we can solve the maximum flow problem on this graph to extract high-quality suppliers with unfair visibility in recommendation lists. Finally, we can reconstruct the final recommendation lists by adding high-quality items from those suppliers according to each user's preferences. 

Similar settings can also be considered on FairMatch for improving user fairness \cite{mansoury2020a}. Considering the job recommendation domain where the task is recommending jobs to users, FairMatch can be formulated to fairly distribute "good" jobs (e.g. highly-paying jobs) to each group of users based on sensitive attributes (e.g. men and women). We consider these scenarios in our future work.

\textbf{Flexibility.} %FairMatch is flexible to be used for various purposes. %As explaoined in generalization part, FairMatch can be used for improving supplier-side fairness by modifying graph construction, weight computation, and recommendation reconstruction.
Another potential interesting improvement on FairMatch is taking into account the item ranking in final recommendation lists. In this paper, we aimed at creating final recommendation lists to include high-quality items with low visibility and we measured it in terms of precision. However, FairMatch allows to consider creating fair ranked lists by modifying the last step (recommendation construction). To do this, given extracted items from step 3 and top $n$ recommendation lists from standard recommendation algorithm, the goal is to find the fair position for extracted items in the top $n$ recommendation.

Finally, weight computation at step 2 also provides flexibility in optimizing FairMatch to capture some other aspects. For instance, considering the popularity of items for computing weights of edges on recommendation bipartite graph may help to further control popularity bias in recommender systems. 

%\todo[inline]{Another piece of future work might be to consider using this algorithm for user fairness. The setting would be one where you have ``good'' items (highly-paying jobs for example) and you want to make sure that they are fairly distributed to users in different groups (say men and women). I think you could switch around the problem formulation to make this work.}

%\textbf{Efficiency.} FairMatch does not require computationally expensive processes. Analogous to state-of-the-art techniques,  

\section{Conclusion}

In this paper, we proposed a graph-based approach, FairMatch, for improving the aggregate diversity of recommender systems. FairMatch is a post-processing technique that works on the top of any recommendation algorithm. In other words, it re-ranks the output from the standard recommendation algorithm such that it improves the aggregate diversity of final recommendation lists, while it maintains the accuracy of recommendations. Experimental results on two publicly available datasets showed that the FairMatch algorithm outperforms several state-of-the-art methods in improving aggregate diversity. One of the limitations of our work is that our algorithm does not leverage the information about the popularity of items in rating data. We believe this information could play an important role in further improving aggregate diversity of the final recommendation lists because usually algorithms are biased toward popular items \cite{himan2019b} and tackling this bias could increase the number of distinct recommended items, hence higher aggregate diversity. We intend to investigate this limitation in future work.
%
% The next two lines define the bibliography style to be used, and the bibliography file.
%\bibliographystyle{ACM-Reference-Format}
%\bibliography{sample-base}

\bibliographystyle{ACM-Reference-Format}
\balance
\bibliography{ref}

\end{document}

%% file: t_ep.tex
\captionsetup[table]{skip=4pt}
\begin{table*}[t]
%\footnotesize
\small
\centering
\setlength{\tabcolsep}{3pt}
\captionof{table}{Comparison of different post-processing techniques on Epinions dataset  for recommendation list of size 10.} \label{tab:ep}
\begin{tabular}{llrrrrrrrrrrrrrr}
\toprule
 \multirow{2}{*}{algorithms} & \multirow{2}{*}{baselines} & \multicolumn{4}{c}{$t=20$} & & \multicolumn{4}{c}{$t=50$} & & \multicolumn{4}{c}{$t=100$} \\\cline{3-6}\cline{8-11}\cline{13-16}
 
 & & P@10 & C@10 & G@10 & E@10 & & P@10 & C@10 & G@10 & E@10 & & P@10 & C@10 & G@10 & E@10 \\
 \bottomrule
 
 \multirow{6}{*}{\algname{ListRankMF}}   & Standard & 
    0.015 & 24.4\% & 0.937 & 4.35 && 
    0.015 & 24.4\% & 0.937 & 4.35 && 
    0.015 & 24.4\% & 0.937 & 4.35 \\
 & Random
    & 0.010 & 33.1\% & 0.882 & 5.23  
    && 0.006 & 45.0\% & 0.869 & 5.97 
    && 0.004 & 53.5\% & 0.856 & 6.39 \\
 & Reverse
    & 0.005 & 37.3\% & 0.839 & 5.69 
    && 0.003 & 51.9\% & 0.867 & 5.58 
    && 0.002 & 60.7\% & 0.814 & 6.70 \\
 & FA*IR 
    & 0.013 & 28.8\% & 0.917 & 4.56 % k20c20p0.75alpha0.1
    && 0.009 & 30.4\% & 0.934 & 5.01 % k50c10p0.95alpha0.1 
    && 0.008 & 33.8\% & 0.945 & 5.01 \\ %k100c20p0.95alpha0.1
% & xQuAD
%    &  &  &  &   
%    &&  &  &  &  
%    &&  &  &  &  \\
 & DM
    & 0.014 & 36.9\% & 0.907 & 4.45 % uniformA1-mu1
    && 0.011 & 56.6\% & 0.748 & 5.69 % uniformA9-mu0.01
    && 0.010 & 69.1\% & \textbf{0.680} & 6.07 \\ % uniformA9-mu0.01
 & FairMatch
    & 0.014 & 38.0\% & 0.884 & 4.72 % d0.0-r1.0 -0.5320-0.09632 - 0.0216
    && 0.010 & \textbf{61.4\%} & 0.789 & 5.96 % d0.0-r1.0 - 0.00010 - 0.0123 - 0.01387
    && 0.008 & \textbf{77.7\%} & 0.720 & \textbf{6.53} \\ %d0.0-r1.0 - 0.00053 - 0.0055 - 0.0078
 
 \hline
 
 \multirow{6}{*}{\algname{UserKNN}}   & Standard & 
    0.045 & 46.4\% & 0.925 & 5.19 && 
    0.045 & 46.4\% & 0.925 & 5.19 && 
    0.045 & 46.4\% & 0.925 & 5.19 \\
 & Random
    & 0.035 & 53.6\% & 0.896 & 5.63  
    && 0.023 & 65.7\% & 0.875 & 6.26 
    && 0.016 & 75.6\% & 0.831 & 6.73 \\
 & Reverse
    & 0.025 & 58.2\% & 0.870 & 5.96  
    && 0.013 & 73.3\% & 0.825 & 6.70 
    && 0.008 & 83.1\% & 0.753 & 7.20 \\
 & FA*IR
    & 0.044 & 61.1\% & 0.868 & 5.62 % k20c30p0.95alpha0.15
    && 0.038 & 65.0\% & 0.865 & 6.22 % k50c30p0.95alpha0.1 
    && 0.030 & 65.4\% & 0.867 & 6.41 \\ % k100c30p0.95alpha0.1
% & xQuAD
%    &  &  &  &   
%    &&  &  &  &  
%    &&  &  &  &  \\
 & DM
    & 0.044 & 64.1\% & \textbf{0.850} & 5.65 % uniformA9-mu1
    && 0.041 & 84.3\% & \textbf{0.732} & 6.40 % uniformA9-mu0.5
    && 0.037 & 95.4\% & 0.529 & 7.18 \\ % uniformA9-mu1
 & FairMatch
    & 0.044 & \textbf{67.0\%} & 0.853 & 5.72 % d0.25-r0.75 - 0.0006 - 0.4505 - 0.04464
    && 0.038 & \textbf{90.8\%} & \textbf{0.732} & 6.67 % d0.0-r1.0 - 0.000005 - 0.98280 - 0.02214
    && 0.029 & \textbf{98.1\%} & 0.580  & \textbf{7.41} \\ % d0.0-r1.0
 
 \bottomrule
\end{tabular}
\end{table*}

%% file: t_ml1m.tex
\captionsetup[table]{skip=4pt}
\begin{table*}[t]
%\footnotesize
\small
\centering
\setlength{\tabcolsep}{3pt}
\captionof{table}{Comparison of different post-processing techniques on MovieLens dataset  for recommendation list of size 10.} \label{tab:ml}
\begin{tabular}{llrrrrrrrrrrrrrr}
\toprule
 \multirow{2}{*}{algorithms} & \multirow{2}{*}{baselines} & \multicolumn{4}{c}{$t=20$} & & \multicolumn{4}{c}{$t=50$} & & \multicolumn{4}{c}{$t=100$} \\\cline{3-6}\cline{8-11}\cline{13-16}
 
 & & P@10 & C@10 & G@10 & E@10 & & P@10 & C@10 & G@10 & E@10 & & P@10 & C@10 & G@10 & E@10 \\
 \bottomrule
 
 \multirow{6}{*}{\algname{ListRankMF}}   & Standard & 
    0.152 & 14.0\% & 0.916 & 4.13 && 
    0.152 & 14.0\% & 0.916 & 4.13 && 
    0.152 & 14.0\% & 0.916 & 4.13 \\
 & Random
    & 0.124 & 17.5\% & 0.861 & 4.76  
    && 0.089 & 24.6\% & 0.834 & 5.48 
    && 0.066 & 32.2\% & 0.809 & 5.97 \\
 & Reverse
    & 0.097 & 19.0\% & 0.831 & 4.97 
    && 0.055 & 28.4\% & 0.786 & 5.73 
    && 0.037 & 37.9\% & 0.757 & 6.20 \\
 & FA*IR 
    & 0.143 & 14.2\% & 0.907 & 4.26 %k20c10p0.75alpha0.05  
    && 0.136 & 14.3\% & 0.937 & 4.34 % k50c10p0.75alpha0.1
    && 0.128 & 16.5\% & 0.949 & 4.41 \\ %k100c20p0.75alpha0.05
 & DM
    & 0.148 & 18.7\% & \textbf{0.850} & 4.41 % uniformA5-mu1 
    && 0.138 & 28.4\% & \textbf{0.801} & 4.76 % uniformA5-mu1
    && 0.130 & 38.1\% & \textbf{0.764} & 5.02 \\ % uniformA5-mu1
 & FairMatch
    & 0.149 & 19.4\% & 0.870 & 4.40 % d0.5-r0.5 - 0.0107 - 0.000506 - 0.42077
    && 0.138 & \textbf{30.0\%} & 0.836 & \textbf{4.90} % d0.25-r0.75 - 0.00265 - 0.0049 - 0.04794
    && 0.130 & 40.2\% & 0.834  & 5.10 \\ % d0.0-r1.0 - 0.01798 - 0.000189 - 0.051140
 
 \hline
 
 \multirow{6}{*}{\algname{UserKNN}}   & Standard & 
    0.196 & 10.7\% & 0.884 & 4.37 && 
    0.196 & 10.7\% & 0.884 & 4.37 && 
    0.196 & 10.7\% & 0.884 & 4.37 \\
 & Random
    & 0.163 & 12.3\% & 0.836 & 4.73  
    && 0.120 & 15.9\% & 0.805 & 5.29 
    && 0.094 & 19.4\% & 0.780 & 5.71 \\
 & Reverse
    & 0.130 & 13.4\% & 0.791 & 4.99  
    && 0.082 & 17.7\% & 0.726 & 5.63 
    && 0.058 & 22.2\% & 0.703 & 6.01 \\
 & FA*IR
    & 0.192 & 11.3\% & 0.855 & 4.60 % k20c10p0.5alpha0.15
    && 0.181 & 12.3\% & 0.869 & 4.88 % k50c10p0.75alpha0.1
    && 0.168 & 18.0\% & 0.858 & 5.19 \\ % k100c30p0.75alpha0.05
% & xQuAD
%    &  &  &  &   
%    &&  &  &  &  
%    &&  &  &  &  \\
 & DM
    & 0.192 & 13.8\% & \textbf{0.835} & 4.63 % uniformA5-mu1 
    && 0.184 & 19.2\% & \textbf{0.800} & \textbf{4.98} % uniformA5-mu1
    && 0.180 & 25.0\% & 0.780 & 5.21 \\ % uniformA5-mu1
 & FairMatch
    & 0.193 & 13.9\% & 0.863 & 4.48 % d0.25-r0.75 - 0.3966 - 0.00003 - 0.00001
    && 0.184 & 18.6\% & 0.872 & 4.69 % d0.0-r1.0 - 0.27231 - 0.000003 - 0.00001
    && 0.170 & 23.6\% & 0.850  & 5.05 \\ % d0.0-r1.0 - 0.1646 - 0.02774 - 0.287564
 
 \bottomrule
\end{tabular}
\end{table*}